\def\dist{\mathop{\rm dist}\nolimits}

\newcommand{\ba}{\boldsymbol{a}}
\newcommand{\bb}{\boldsymbol{b}}

\newcommand{\bu}{\boldsymbol{u}}
\newcommand{\bv}{\boldsymbol{v}}

\newcommand{\bx}{\boldsymbol{x}}
\newcommand{\by}{\boldsymbol{y}}

\newcommand{\bU}{\boldsymbol{U}}
\newcommand{\bV}{\boldsymbol{V}}
\newcommand{\bW}{\boldsymbol{W}}
\newcommand{\bX}{\boldsymbol{X}}
\newcommand{\bY}{\boldsymbol{Y}}

\newcommand{\convexcluster}{{\sc convexcluster}}
\newcommand{\hclust}{{\sc hclust}}
\newcommand{\admixture}{{\sc admixture}}
\newcommand{\clusterpath}{{\sc clusterpath}}
\newcommand{\structure}{{\sc structure}}
\newcommand{\mi}{{\sc mi}}
\newcommand{\eigenstrat}{{\sc eigenstrat}}

\documentclass[titlepage,10pt]{article}
\usepackage[authoryear,square]{natbib}
\usepackage{graphicx}
\usepackage{pdfpages}
\usepackage{amsmath}
\usepackage{rotating}
\usepackage{listings}
\usepackage{setspace}
\usepackage[bottom=1in,right=1in,top=1.5in,left=1in]{geometry}
\usepackage{indentfirst}
\usepackage{url}
\usepackage{color}

\begin{document}

\title{Convex Clustering: An Attractive Alternative to Hierarchical Clustering}
\date{}
\author{
Gary K. Chen\thanks{Department of Preventive Medicine (Biostatistics Division), University of Southern California, CA} \and 
Eric C. Chi\thanks{Department of Electrical and Computer Engineering, Rice University, TX} \and 
John Michael O. Ranola\thanks{Department of Statistics, University of Washington, WA} \and 
Kenneth Lange\thanks{Departments of Biomathematics, Human Genetics, and Statistics, University of Los Angeles, CA}}

\maketitle

\begin{abstract}

The primary goal in cluster analysis is to discover natural groupings of objects. The field of cluster analysis is crowded with diverse methods that make special assumptions about data and address different scientific aims. Despite its shortcomings in accuracy, hierarchical clustering is the dominant clustering method in bioinformatics. Biologists find the trees constructed by hierarchical clustering visually appealing and in tune with their evolutionary perspective. Hierarchical clustering operates on multiple scales simultaneously. This is essential, for instance, in transcriptome data where one may be interested in making qualitative inferences about how lower-order relationships like gene modules lead to higher-order relationships like pathways or biological processes. The recently developed method of convex clustering preserves the visual appeal of hierarchical clustering while ameliorating its propensity to make false inferences in the presence of outliers and noise. The current paper exploits the proximal distance principle to construct a novel algorithm for solving the convex clustering problem. The solution paths generated by convex clustering reveal relationships between clusters that are hidden by static methods such as k-means clustering. Our convex clustering software separates parameters, accommodates missing data, and supports prior information on relationships. The software is implemented on ATI and nVidia graphics processing units (GPUs) for maximal speed. Several biological examples illustrate the strengths of convex clustering and the ability of the proximal distance algorithm to handle high-dimensional problems.

\end{abstract}

\section{Author Summary}

Pattern discovery is one of the most important goals of data-driven research. In the biological sciences hierarchical clustering  has achieved a position of pre-eminence due to its ability to capture multiple levels of data granularity. Hierarchical clustering's visual displays of phylogenetic trees and gene-expression modules are indeed seductive.  Despite its merits, hierarchical clustering is greedy by nature and often produces spurious clusters, particularly in the presence of substantial noise. This paper presents a relatively new alternative to hierarchical clustering known as convex clustering. Although convex clustering is more computationally demanding, it enjoys several advantages over hierarchical clustering and other traditional methods of clustering. Convex clustering delivers a uniquely defined clustering path that partially obviates the need for choosing an optimal number of clusters. Along the path small clusters gradually coalesce to form larger clusters. Clustering can be guided by external information through appropriately defined similarity weights.  The current paper introduces a new algorithm for solving the convex clustering problem and applies it a variety of biological datasets. Comparisons to hierarchical clustering demonstrate the superior robustness of convex clustering. Our genetics examples include inference of the demographic history of 52 populations across the world, a more detailed analysis of European demography, and a re-analysis of a well-known  breast cancer expression dataset. Our new algorithm is a particular example of a class of MM algorithms known as proximal distance algorithms. The proximal distance convex clustering algorithm is inherently parallel and readily maps to modern many-core devices such as graphics processing units (GPUs). Our freely available software, \convexcluster, exploits OpenCL routines that ensure compatibility across a variety of hardware environments.

\section{Introduction}

Pattern discovery is one of the primary goals of bioinformatics. Cluster analysis is a broad term for a variety of exploratory methods that reveal patterns based on similarities between data points. Well-known methods such as $k$-means invoke a fixed number of clusters. In complex biological data, the number of clusters is unknown in advance, and it is appealing to vary the number of clusters simultaneously with cluster assignment. Hierarchical clustering has been particularly helpful in understanding cluster granularity in gene-expression studies and other applications. In addition to producing easily visualized and interpretable results, hierarchical clustering is simple to implement and computationally quick. These are legitimate advantages, but they do not compensate for hierarchical clustering's instability to small data perturbations.

All principled methods of clustering attempt to decrease some criterion. Hierarchical clustering constructs a dendrogram by fusing or dividing observations (features). Fusion is referred to as agglomerative clustering and splitting as divisive clustering. Because of the greedy nature of the choices in hierarchical clustering, it returns clusters that are only locally optimal with respect to the underlying criterion. Solutions may vary depending on how the algorithm is initialized. To improve the chances of reaching a global minimum, multiple algorithm initializations must be tried. Even then there is no guarantee of optimality. A potentially greater handicap is that small perturbations in the data can lead to large changes in hierarchical clustering assignments. This propensity makes hierarchical clustering sensitive to outliers and promotes the formation of spurious clusters. In combination, the presence of local minima and the sensitivity to outliers lead to irreproducible results.

Although these objections are serious, a complete reformulation of hierarchical clustering is unnecessary. Recently \cite{lindsten11} and \cite{hocking11} introduced convex clustering based on minimizing a penalized sum of squares. The strict convexity and coercivity of their criterion guarantees a unique clustering path. The penalty term in convex clustering criterion accommodates prior information through nonuniform weights on data pairs. The solution paths of convex clustering retain the straightforward interpretability of hierarchical clustering while ameliorating its sensitivity to outliers and tendency to get trapped by local minima.

Despite the promise of convex clustering, there are two obstacles that stand in its way of becoming a practical tool in bioinformatics. The first is the challenge of large-scale problems. Current algorithms are computationally intensive and scale poorly on high-dimensional  problems. A second obstacle is the minimal guidance currently available on how to choose penalty weights. \cite{hocking11} 
suggest some rules of thumb but offer little detailed advice. In our experience, the quality of the clustering path depends critically on well-designed weights. To address these issues, the current paper describes a fast new algorithm and a corresponding software implementation, \convexcluster. Our advice on strategies for choosing penalty weights is grounded in a few practical biological examples. These examples support our conviction that convex clustering can be more nuanced than hierarchical clustering. Our examples include Fisher's Iris data from discriminant analysis, ethnicity clustering based on microsatellite genotypes from the Human Genome Diversity Project and SNP genotypes from the POPRES project, and breast cancer subtype classification via microarrays. In the POPRES data, we first reduce the genotypes to principal components and then use these to cluster.  The paths computed under convex clustering  expose features of the data hidden to less sophisticated clustering methods. The potential for understanding human evolution and history alone justify wider adoption of convex clustering.

\section{Methods}

Assume that there are $n$ cases and $p$ predictors. The more vivid language of graph theory
speaks of nodes rather than cases and edges rather than pairs of cases. To implement convex clustering,  \cite{lindsten11} suggest minimizing the penalized loss function
\begin{eqnarray}
f_\mu(\bU) &=& \frac{1}{2}\sum_{i=1}^n ||\bx_i-\bu_i||^2 + \mu \sum_{i<j}w_{ij}||\bu_i-\bu_j||
\label{eq_lindsten}
\end{eqnarray}
relying on Euclidean norms. Here the column vector $\bx_i$ of the matrix $\bX$ records the predictors for case $i$, the column $\bu_i$ of the matrix $\bU$ denotes the cluster center assigned to case $i$, $\mu \ge 0$ tunes the strength of the penalty, and $w_{ij} \ge 0$ weights the contribution of the case pair $(i,j)$ to the penalty.
The objective function $f_\mu(\bU)$ treats the predictors symmetrically. If these range over widely
varying scales, it is prudent to standardize each predictor to have mean 0 and variance 1.

Because the objective function $f_\mu(\bU)$ is strictly convex and coercive, a unique minimum point exists for each value of $\mu$. When $\mu=0$, the values $\bu_i=\bx_i$ minimize $f_\mu(\bU)$, and there are as many clusters as cases. If the underlying graph is connected, then as $\mu$ increases, cluster centers coalesce until all centers merge into a single cluster with all $\bu_i=\bar{\bx}$, the average of the data points 
$\bx_i$. Although fission events as well as fusion events can in principle occur along the solution path, following the path as $\mu$ increases typically reveals a hierarchical structure among the clusters. The weights encode prior information that guides clustering. Setting some of the weights equal to 0 reduces the computational load of minimizing $f_\mu(\bU)$ in the proximal distance algorithm introduced next.

\subsection{The Proximal Distance Algorithm}

The proximal distance principle is a new way of attacking constrained optimization 
problems \citep{lange2014congress}. The principle is capable of enforcing parsimony in parameter estimation while avoiding the shrinkage incurred by convex penalties such as the lasso. Shrinkage leads to imperfect model selection in addition to poor parameter 
estimates. The proximal distance principle seeks to minimize a function $h(\by)$, possibly nonsmooth, subject to $\by \in C$, where $C$ is a closed set, not necessarily convex.
The set $C$ encodes constraints such as sparsity. In the exact penalty method of Clarke \citep{BorweinLewis06ConvexAnalysisBook,clarke1990optimization,demyanov2010nonlinear}, 
this constrained problem is replaced by the unconstrained problem of minimizing $h(\by)+\rho \dist(\by,C)$, where $\dist(\by,C)$ denotes the Euclidean distance from $\by$ to $C$. Note that $\dist(\by,C)=0$ is a necessary and sufficient condition for $\by \in C$. If $\rho$ is chosen large enough, say bigger than a Lipschitz constant for $h(\by)$, then the minima of the two problems coincide.

How does convex clustering fit in this abstract framework? Although the objective 
function $f_\mu(\bU)$ is certainly nonsmooth, there are no constraints in sight.
The strategy of parameter splitting introduces constraints to simplify the
objective function. Since least squares problems are routine, the penalty 
terms constitute the intractable part of the objective function $f_\mu(\bU)$. 
One can simplify the term $\|\bu_i-\bu_j\|$ by replacing the vector difference 
$\bu_i-\bu_j$ by the single vector $\bv_{ij}$ and imposing the constraint 
$\bv_{ij} = \bu_i-\bu_j$. Parameter splitting therefore leads to the revised 
objective function
\begin{eqnarray}
g_\mu(\bU,\bV) &=& \frac{1}{2}\sum_{i=1}^n||\bx_i-\bu_i||^2 + \mu \sum_{i<j}w_{ij}||\bv_{ij}||
\label{eq_proxdist}
\end{eqnarray}
with a simpler loss, an expanded set of parameters, and a linear constraint
set $C$ encapsulating the pairwise constraints $\bv_{ij} = \bu_i-\bu_j$.

The proximal distance method undertakes minimization of $h(\by)+\rho \dist(\by,C)$
by a combination of approximation, the MM (majorization-minimization) principle
\citep{borg2005modern,heiser1995convergent,HunterLange04MMTutorial,Lange2000opttransfer,WuLange10EMMM}, 
and an appeal to a combination of set projection \citep{Deutsch02Book} and proximal mapping 
\citep{parikh2013proximal}. The latter 
operations have been intensely studied for years and implemented
in a host of special cases. Thus, the proximal distance principle encourages
highly modular solutions to difficult optimization problems.  Furthermore, 
most proximal distance algorithms benefit from parallelization.

Let us consider each of the ingredients of the proximal distance algorithm in turn,
starting with approximation. The function $\dist(\by,C)$ is nonsmooth
even when $C$ is well behaved. For $\epsilon>0$ small, the revised distance 
$\dist_\epsilon(\by,C) =  \sqrt{\dist(\by,C)^2+\epsilon}$
is differentiable and approximates $\dist(\by,C)$ well. The MM principle 
leads to algorithms that systematically decrease the
objective function. In the case of minimizing $f(\by)+\rho \dist(\by,C)$
one can invoke the majorization $\dist(\by,C) \le \|\by-P_C(\by_m)\|$, where $P_C(\by_m)$
is the projection of the current iterate $\by_m$ onto the set $C$. By definition
$\dist(\by_m,C) = \|\by_m-P_C(\by_m)\|$, and $P_C(\by_m)$ is a
closest point in $C$ to the point $\by_m$.  For a closed nonconvex set, there may
be multiple closest points; for a closed convex set there is exactly one.

According to the MM principle, minimizing the surrogate function
\begin{eqnarray}
\frac{1}{2}\sum_{i=1}^n||\bx_i-\bu_i||^2+ \mu \sum_{i<j}w_{ij}||\bv_{ij}||
+\rho \sqrt{\left\|\begin{pmatrix}\bU \\ \bV \end{pmatrix} 
-P_C\begin{pmatrix}\bU_m \\ \bV_m \end{pmatrix} \right\|^2+\epsilon} \label{first_surrogate}
\end{eqnarray}
drives the approximate objective function
\begin{eqnarray*}
\frac{1}{2}\sum_{i=1}^n ||\bx_i-\bu_i||^2+ \mu \sum_{i<j}w_{ij}||\bv_{ij}||
+\rho \sqrt{\dist\left[\begin{pmatrix}\bU \\ \bV \end{pmatrix},C\right]^2+\epsilon}
\end{eqnarray*}
downhill. The surrogate function (\ref{first_surrogate}) is still too complicated for
our purposes. The remedy is another round of majorization. This time the majorization
\begin{eqnarray}
\sqrt{t+\epsilon} & \le & \sqrt{t_m+\epsilon}+\frac{1}{2\sqrt{t_m+\epsilon}}(t-t_m)
\label{sqrt_root_majorization}
\end{eqnarray}
comes into play based on the concavity of the function $\sqrt{t+\epsilon}$ for $t \ge 0$.
As required by the MM principle, equality holds in the majorization (\ref{sqrt_root_majorization})
when $t=t_m$. Applying this majorization to the surrogate function (\ref{first_surrogate})
yields the new surrogate
\begin{eqnarray}
h[(\bU,\bV) \mid (\bU_m,\bV_m)] & = & \frac{1}{2}\sum_{i=1}^n||\bx_i-\bu_i||^2+ \mu \sum_{i<j}w_{ij}||\bv_{ij}||
+\frac{\rho}{2d_m}\left\|\begin{pmatrix}\bU \\ \bV \end{pmatrix} 
-P_C\begin{pmatrix}\bU_m \\ \bV_m \end{pmatrix} \right\|^2 \label{second_surrogate} \\
d_m & = & \sqrt{\left\|\begin{pmatrix}\bU_m \\ \bV_m \end{pmatrix} 
-P_C\begin{pmatrix}\bU_m \\ \bV_m \end{pmatrix} \right\|^2+\epsilon}
\nonumber
\end{eqnarray}
up to an irrelevant constant. The surrogate function (\ref{second_surrogate}) resulting from 
these maneuvers separates all of the vectors $\bu_i$ and $\bv_{ij}$. One can explicitly solve for the updates
\begin{eqnarray*}
\bu_{n+1,i} & = & \frac{d_m}{d_m+\rho}\bx_i+\frac{\rho}{d_m+\rho}\ba_{ni},
\end{eqnarray*}
where $\ba_{n,i}$ is the part of the projection pertaining to $\bu_i$. The update of $\bv_{ij}$
involves shrinkage. Let $\bb_{n,ij}$ denote the part of the projection pertaining to
$\bv_{ij}$. Standard arguments from convex calculus \citep{Lange2012optimization}
show that the minimum of $\mu w_{ij}\|\bv_{ij}\|+\frac{\rho}{2d_m}\|\bv_{ij}-\bb_{n,ij}\|^2$ is achieved by
\begin{eqnarray*}
\bv_{n+1,ij} & = & \max\left\{\left(1-\frac{\mu w_{ij} d_m}{\rho\|\bb_{n,ij}\|}\right),0\right\}\bb_{n,ij}.
\end{eqnarray*}
In the exceptional case $\bb_{n,ij} = {\bf 0}$,
the solution $\bv_{n+1,ij} = {\bf 0}$ is clear from inspection of the $\bv_{ij}$ 
criterion. Both of these solution maps fall under the heading of proximal operators,
hence, the name proximal distance algorithm.

If a weight $w_{ij}=0$, then it is computationally foolish to introduce
a difference vector $\bv_{ij}$. In many applications, the weight matrix
$\bW=(w_{ij})$ may be sparse. Given this lack of symmetry, one cannot expect to project analytically onto the constraint space. We
now discuss a block descent algorithm for projection.  Let $E$ denote the
set of edges $\{i,j\}$ with positive weights $w_{ij}= w_{ji}$. Divide the
neighborhood $N_i$ of a node $i$ into left and right neighborhoods $L_i = \{j<i: w_{ji}>0\}$
and $R_i = \{j>i: w_{ij}>0\}$. Clearly $N_i = L_i \cup R_i$, and $E = \cup_{i=1}^n  N_i$. 
Projection minimizes the criterion
\begin{eqnarray*}
\frac{1}{2}\sum_{i=1}^n \|\bu_i-\tilde{\bu}_i\|^2 
+\frac{1}{2}\sum_{\{i,j\} \in E}\|\bu_i-\bu_j-\tilde{\bv}_{ij}\|^2
\end{eqnarray*}
for $\tilde{\bU}$ and $\tilde{\bV}$ given. It is unclear how to massage the stationarity equations
\begin{eqnarray*}
{\bf 0} & = & \bu_i-\tilde{\bu}_i+\sum_{j \in R_i}(\bu_i-\bu_j-\tilde{\bv}_{ij})
-\sum_{j \in L_i}(\bu_j-\bu_i-\tilde{\bv}_{ji})
\end{eqnarray*}
into a solvable form. However, the block updates
\begin{eqnarray*}
\bu_i & = & \frac{1}{1+|N_i|}\left(\tilde{\bu}_i+\sum_{j \in R_i}\tilde{\bv}_{ij}
-\sum_{j \in L_i}\tilde{\bv}_{ji}+\sum_{j \in N_i} \bu_j\right)
\end{eqnarray*}
are available. Here $|N_i|$ denotes the cardinality of $N_i$.  One cycle of the
block descent algorithm updates $\bu_1$ through $\bu_n$ sequentially. This 
cycle is repeated until all of the vectors $\bu_i$ stabilize. Once convergence
is achieved, one sets $\bv_{ij}=\bu_i-\bu_j$ for the relevant pairs.

\subsection{Missing Data}

The convex function (\ref{eq_proxdist}) assumes no missing entries in the data matrix $\bX$. It is straightforward to accommodate missing data by another round of majorization. Suppose $\Gamma$ is the set of ordered index pairs $(i,j)$ corresponding to the observed entries $x_{ij}$ of $\bX$.  We now minimize the revised
criterion
\begin{eqnarray}
f_\mu(\bU) &=& \frac{1}{2}\sum_{(i,j) \in \Gamma} (x_{ij}-u_{ij})^2 + \mu \sum_{i<j}w_{ij}||\bu_i-\bu_j|| ,
\label{eq_missing_analog}
\end{eqnarray}
which unfortunately lacks the symmetry of the original problem. To restore the lost symmetry, we
invoke the majorization 
\begin{eqnarray*}
\frac{1}{2}\sum_{(i,j) \in \Gamma} (x_{ij}-u_{ij})^2 & \le &
\frac{1}{2}\sum_{(i,j) \in \Gamma} (x_{ij}-u_{ij})^2+\frac{1}{2}\sum_{(i,j) \not\in \Gamma} (u_{mij}-u_{ij})^2,
\end{eqnarray*}
where $u_{mij}$ is a component of $\bU_m$. In essence, the term $(u_{mij}-u_{ij})^2$ majorizes 0.  
If the $n \times p$ matrix $\bY =(y_{ij})$
has entries $y_{ij} = x_{ij}$ for $(i,j) \in \Gamma$ and $y_{ij}=u_{mij}$  for $(i,j) \not\in \Gamma$,
then in the minimization step of the proximal distance algorithm, we simply minimize the surrogate function
\begin{eqnarray}
g_\mu(\bU,\bV) &=& \frac{1}{2}\sum_{i=1}^n||\by_i-\bu_i||^2 + \mu \sum_{i<j}w_{ij}||\bv_{ij}||
\label{eq_proxdist}
\end{eqnarray}
The rest of the proximal distance algorithm remains the same

\subsection{Calibration of Weights}

The pairwise weight $w_{ij}=w_{ji}$ introduced in the penalty term of equation (\ref{eq_lindsten}) determines the importance of similarity between nodes $i$ and $j$. Two principles guide our choice of weights. First, the weight $w_{ij}$ should be inversely proportional to the distance between the $i$th and $j$th points.
This inverse relationship accords with intuition. As $w_{ij}$ increases, the pressure for the $i$th and $j$th centroids to coalesce increases.  If the weights $w_{ij}$ are correlated with the similarity of the feature vectors $\bx_i$ and $\bx_j$, then the pressure for their centroids to merge is especially great. Second, the weight matrix $\bW$ should be sparse. Despite the fact that small positive weights and zero weights lead to similar clustering paths, the computational advantages of zero weights cannot be ignored. 

These observations prompt the following choice of weights. To maintain computational efficiency, it is helpful to focus on the $k$ nearest neighbors of each node. We define the distance $d_{ij}$ between two nodes $i$ and $j$ by the Euclidean norm $||\bx_i - \bx_j ||$ and write $i \sim_k j$ if $j$ occurs
among the $k$ nearest neighbors of $i$ or vice versa. Based on these considerations the weights
\begin{eqnarray}
w_{ij} & = & 1_{\{i \sim_k j\}} e^{-\phi d_{ij}^2}
\label{eq_weighting}
\end{eqnarray}
are reasonable, where $1_{\{i \sim_k j\}}$ is the indicator function of the event $\{i \sim_k j\}$ and $\phi \ge 0$ is a tuning constant. The case $\phi=0$ corresponds to uniform weights between nearest neighbors. When $\phi$ is positive, $w_{ij}$ strictly decreases as a function of $d_{ij}$. The relation $i \sim_k j$ partitions the nodes into disjoint equivalence classes. Complete coalescence of the nodes occurs as $\mu$ increases if and only if there is a single equivalence class and the graph is connected. Using squared distances $d_{ij}^2$ rather than distances $d_{ij}$ induces more aggressive coalescence of nearby points and slower coalescence of distant points. In practice we normalize weights so that they sum to 1. This harmless tactic is equivalent to rescaling $\mu$.

This generic framework was proposed by \cite{hocking11}. We now discuss a strategy for leveraging additional information.
When expert knowledge on the relationships among nodes is available and can be quantified, incorporating such knowledge may improve the clustering path. This must be done delicately so that  prior information does not overwhelm observed data. In two of our examples, we integrate both genetic and geographic proximity measures in the weights $w_{ij}$. If $\bx_i$ and $\by_i$ store the
genotypes and GPS (global positioning system) coordinates on subject $i$, respectively, then the weighted average
\begin{eqnarray}
d_{ij} & = & \alpha \|\bx_i-\bx_j\| + (1-\alpha) \|\by_i-\by_j\|, \quad \quad \alpha \in (0,1),
\label{composite_distance}
\end{eqnarray} 
serves as a composite distance helpful in clustering subjects. Observe that the components of the difference $\by_i-\by_j$ must be computed in modulo arithmetic. Given a proper choice of the scaling constant $\alpha$, an even better alternative replaces $\|\by_i-\by_j\|$ by the geodesic distance between $i$ and $j$.

\subsection{Evaluation of Clusters}

Our program \convexcluster~minimizes the penalized loss (\ref{eq_proxdist}) for a range of user specified $\mu$ values.  For each $\mu$ the optimized matrix $\bU$ of cluster centers is stored in a temporary file for later construction of the cluster path. To facilitate visualization, \convexcluster~encourages users to project the cluster path onto any two principal components of the original data. The first example of Section \ref{results_section} relies on the classical Iris data of discriminant analysis \citep{fisher1936use}. This dataset contains 150 cases spread over three species. The Iris data can be downloaded from the UCI machine learning repository \citep{BacheLichman2013}. For purposes of comparison, we also evaluated the clusters formed by agglomerative hierarchical clustering. Hierarchical clustering comes in several flavors; we chose UPGMA (Unweighted Pair Group Method with Arithmetic Mean) \citep{sokal58} as implemented in the R function \textit{hclust}. Although \textit{hclust} offers six other options for merging clusters, UPGMA is probably the most reliable in reducing the detrimental effects of outliers since it averages information across all cluster members. UPGMA operates on a matrix of pairwise distances defined between nodes. In our genetics examples, we take these to be the distances defined by equation (\ref{composite_distance}) and used in convex clustering.

\section{Results \label{results_section}}

\subsection{Impact of the Constants $k$ and $\phi$}

\begin{figure}
\begin{center}
\includegraphics[width=7in]{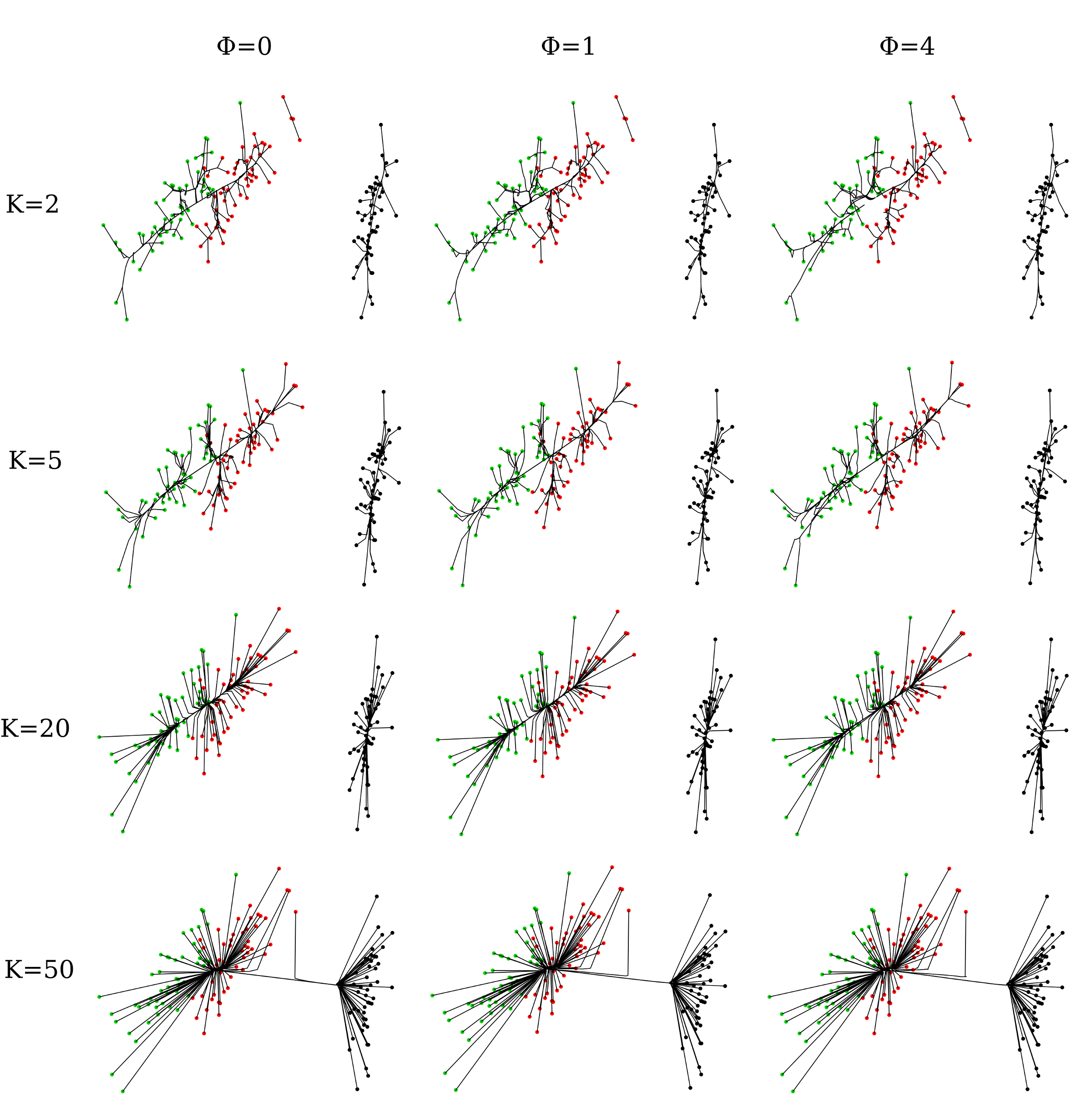}
\caption{\label{fig_iris}Convex clustering of the Iris data. Black, red, and green points denote the species Iris-setosa, Iris-versicolor, and Iris-virginica, respectively. These points are projections of the Iris dataset on the first two principal components (PCs). Lines trace the cluster centers as they traverse the regularization path.}
\end{center}
\end{figure}

To get a sense of the impact of the constants $k$ and $\phi$ on the Iris data, we generated cluster paths for various pairs $(k,\phi)$ . As Figure \ref{fig_iris} illustrates, the number of nearest neighbors $k$ determines the connectivity of the underlying graph. Eventual coalescence only occurs for $k=50$; even then the apparent Iris-Versicolor outlier does not coalesce until very late. All values of $k$ support a clear separation of Iris-Setosa from the other two species Iris-Versicolor and Iris-Virginica.  Separation of Iris-Versicolor and Iris-Virginica into two different groups becomes discernible at $k=20$. Subgroups within each species are evident for $k=5$ and $k=2$. Improved resolution comes at a price; the two small two-member clusters seen in the top right corner of the main Iris-Versicolor cluster never fully coalesce with the main cluster when $k=2$.  The distance tuning constant $\phi$ also exerts subtle influence along each row of Figure \ref{fig_iris}. This influence is more strongly felt for low values of $k$. Examination of the Iris data suggests exploring cluster granularity over a range of $k$ values with $\phi$ set to 0. One can find the minimum $k$ ensuring full connectivity by combining bisection with either breadth-first search or depth-first search \citep{hopcroft1973algorithm}. Once the desired granularity is achieved, $\phi$ can be increased to reveal more subtle details. Note that increasing $\phi$ sends most weights between $k$ nearest neighbors to 0. As previously noted, the proximal distance algorithm takes substantially more iterations to converge for large values of $\phi$.

\subsection{Cluster Accuracy in the Presence of Noise}

Although agglomerative hierarchical clustering is computationally efficient, it tends to create spurious clusters due to its greedy nature. In particular, it can can falter in the face of noisy data. To test this hypothesis, we simulated new data from the Iris data. In creating a dataset, we perturbed each row of the data matrix $\bX$ by adding normal deviates with mean 0 and standard deviation equal to the sample standard deviation $s^2$ of the corresponding predictor multiplied by a constant $c$. We then clustered the data points into three clusters
and counted the number of inconsistencies between cluster labels and species labels. For convex clustering, visual inspection of the converged clustering paths reveals roughly three major clusters for values of $k$ between 5 and 15. With hierarchical clustering, three clusters were constructed by choosing a cut point on the full tree intersecting three branches.  Both methods performed equally well on the original dataset, misclassifying the same 14 Iris-Virginica specimens as Iris-Versicolor for an overall error rate of $14/150=0.093$. One of these errors in possibly a mis-attribution of species; the remaining may represent hybrid plants. Table \ref{table_simulation} summarizes error rates averaged over 100 replicates under the two methods. Examination of the table suggests that convex clustering is indeed more accurate in the face of noise over a wide range of $k$ values.

\subsection{Cluster Accuracy with Missing Values}

We carried out a second simulation study on the Iris data to assess the impact of varying levels of missingness on cluster inference. Because the Iris data includes only four features, simply selecting entries of the data matrix at random can lead to cases retaining no data. To avoid these degeneracies, we randomly selected cases and then a random feature from each case for deletion. Given cases rates of 25\%, 50\%, 75\%, and 100\%, the proportion of missing observations consequently ranged from 5\% to 25\%. Hierarchical clustering with missing data requires that either cases with missing entries be omitted or that missing entries be imputed. We employed the second strategy, filling in missing entries by multiple imputation as implemented in the R package \mi~\citep{gelman2011}. Hierarchical clustering was then applied to the completed data. For convex clustering, we also applied multiple imputation, but for the sole purpose of computing the convex clustering weights. We then applied convex clustering to the original incomplete data under the objective function (\ref{eq_missing_analog}). Accuracy for each method was estimated in the same manner as the previous simulations. The error rates in Table \ref{table_missingness} suggest that convex clustering does indeed outperform hierarchical clustering in the presence of missing data.

\subsection{Inference of Ethnicity}

As genotyping costs have dropped in recent years, it has become straightforward to relate ethnicity to subtle genetic variations. Several software tools are now available for this purpose. For example, the programs \structure~\citep{pritchard00} and \admixture~\citep{alexander2009fast} estimate a subject's admixture proportions across a set of predefined or inferred ancestral populations. \eigenstrat~\citep{price06} employs a handful of principal components to explain ethnic variation. Principal component analysis (PCA) is attractive due to its speed and ease of visualization.  Clustering can also separate people by ethnicity if individuals of mixed ethnicity are omitted. The advantage of convex clustering is that one can follow the dynamic behavior of the relationship clusters along the regularization path.

\subsubsection{World-wide Genetic Diversity}

For a practical demonstration of convex clustering, we now turn to the Human Genome Diversity Project (HGDP). This collaboration makes several datasets publicly available that vary in marker type (SNPs versus microsatellites) and sample size. The HGDP 2002 dataset considered here includes 1,056 individuals from 52 populations genotyped at 377 autosomal microsatellites \citep{rosenberg02}. Care must be taken in analyzing microsatellites since, in contrast to SNPs, they display more alleles and greater levels of polymorphism. Recall that an allele at a microsatellite approximates the number of short tandem repeats of some simple motif.  Because treating microsatellite genotypes as continuous variables is problematic, we encode each microsatellite genotype as a sequence of allele counts. Each count ranges from 0 to 2, and there are as many count variables as alleles.  This encoding yields a revised 2002 dataset with the 377 microsatellite genotypes expanded to 4,682 different attributes.

\begin{figure}
\begin{center}
\includegraphics[width=7in]{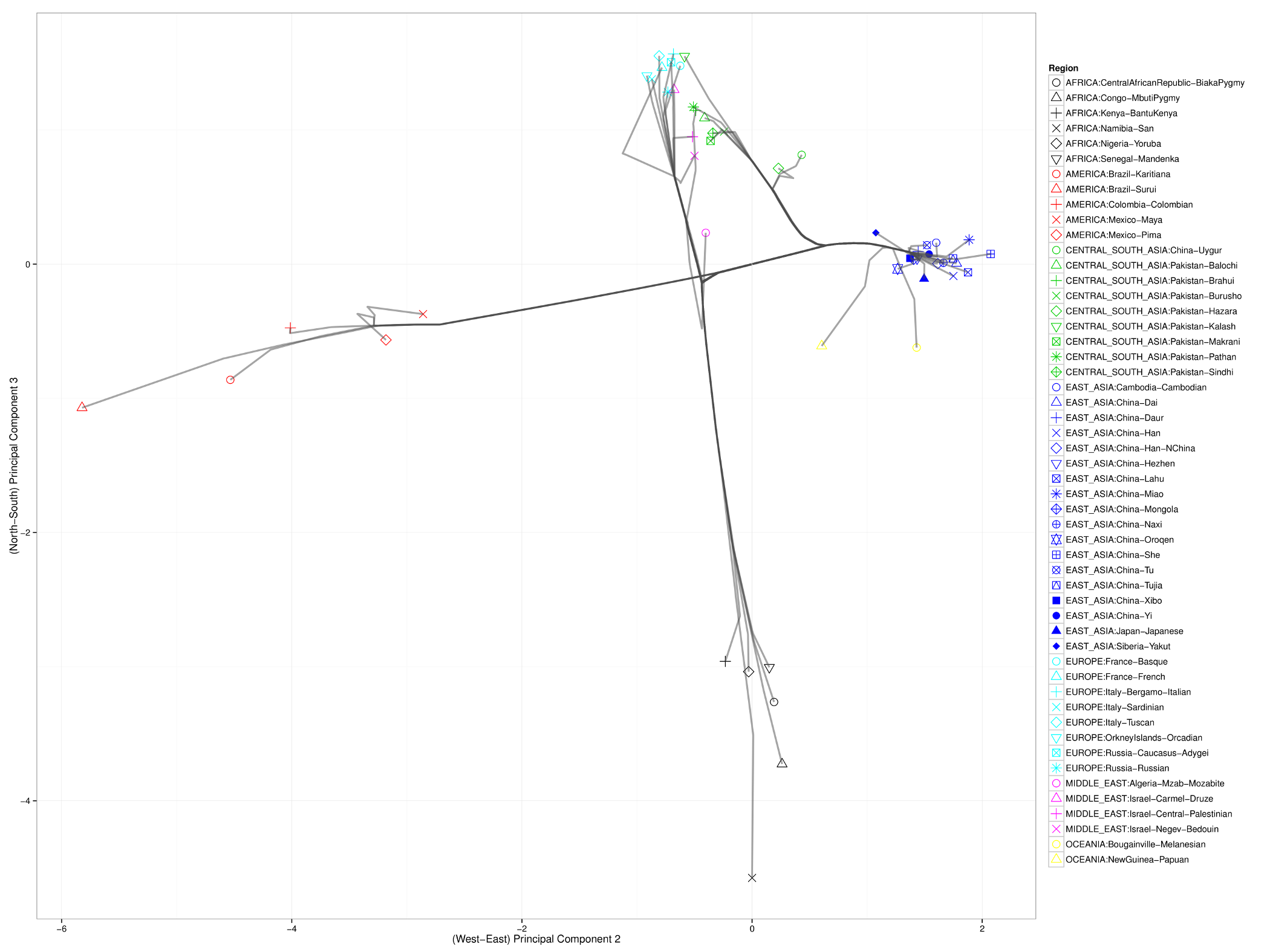}
\caption{\label{fig_convex_hgdp_nozoom}Convex clustering of the HGDP data using a large number  $k$ of nearest neighbors to infer intercontinental connections.}
\end{center}
\end{figure}

\begin{figure}
\begin{center}
\includegraphics[width=7in]{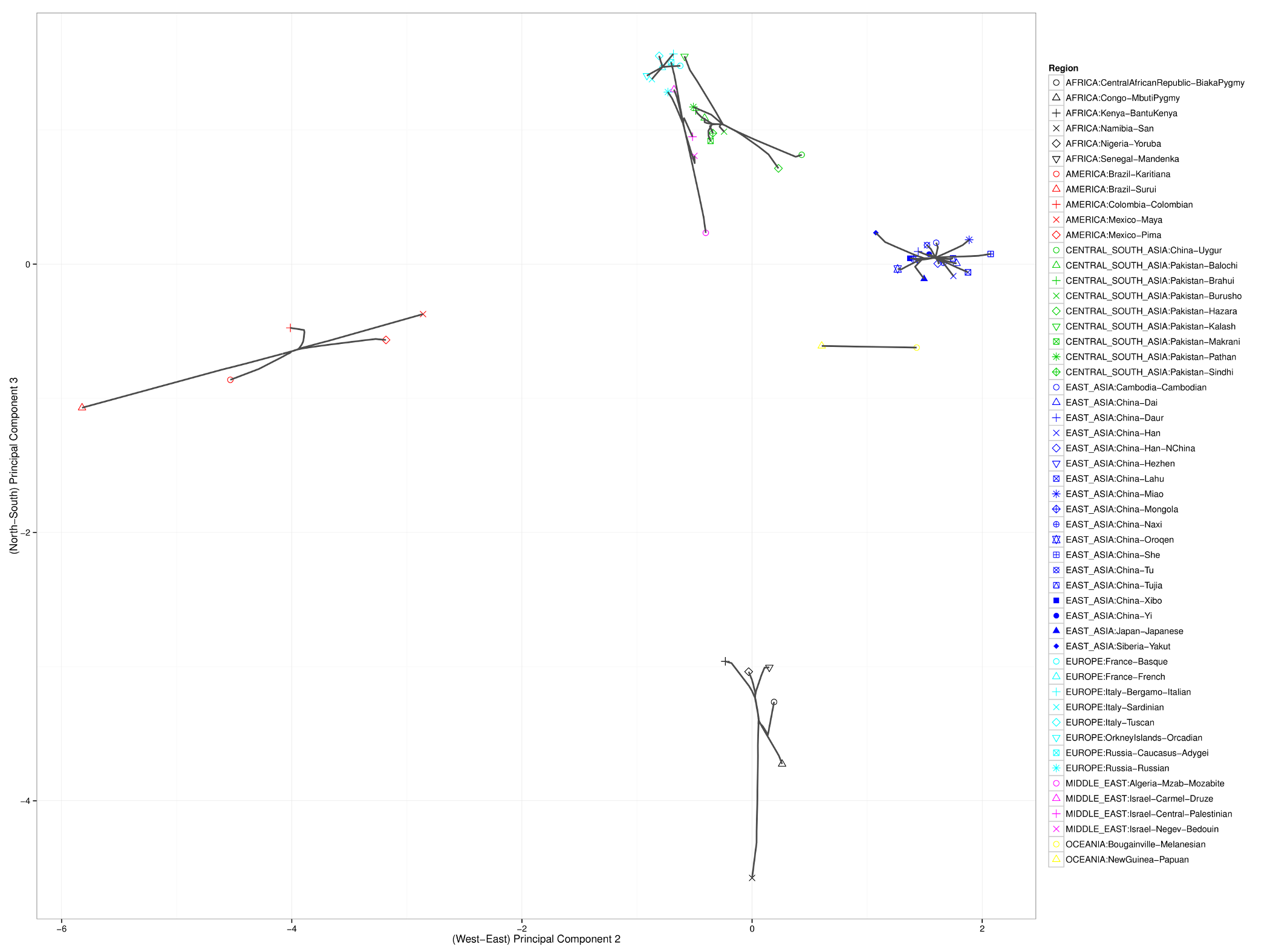}
\caption{\label{fig_convex_hgdp_zoom}Convex clustering of the HGDP data using a small number
$k$ of nearest neighbors to resolve intracontinental connections.}
\end{center}
\end{figure}

As expected, these data exhibit clines in allele frequencies \citep{kittles03}. To take advantage of
the correlation between geographic separation and ethnic similarity, we defined penalty weights $w_{ij}$ according to the composite distance in equation (\ref{composite_distance}) with constant 
$\alpha = 0.5.$  Figure \ref{fig_convex_hgdp_nozoom} plots cluster paths for these data given
the settings $\phi=1$ and $k=4$. With $k=4$ nearest neighbors, we observe broad-scale clustering events that link up the major continental groups. In the north, Europeans fall into a single cluster,
later joined by populations from the Middle East. In the east the Chinese merge into a cluster that
subsequently merges with two Oceania populations from New Guinea. This mega cluster then merges with  various Central Asian populations of predominantly Pakistani origin. In the west five Central/South American populations cluster, and in the south six African populations cluster.  Considering the continental clusters, the first two to coalesce are the American cluster and the Central/East Asian cluster. This accords with known links between East Asians and American Indians, who crossed the Bering strait, possibly multiple times, during the Ice Age.  Figure \ref{fig_convex_hgdp_zoom} depicts finer grained events exposed by setting $k=1$.  Along the western axis, taking $k=1$ is uninformative, but among the  African populations along the southern axis, we observe three major clusters: a two-member cluster representing the two Pygmy sub-groups; a three-member cluster comprising Bantu-speaking peoples from Kenya, Yorubans from Nigeria, and Mandenkas from Senegal; and finally a singleton cluster for the San from Namibia. These results are consistent with a recent phylogenetic study \citep{li08} that found the San to be the most isolated of the African populations, followed by the two Pygmy populations, and finally the three Bantu-language populations. Along the eastern axis, the two Papua New Guinea populations cluster together and do not join the remaining Asian populations. 

\begin{figure}
\begin{center}
\includegraphics[width=7in]{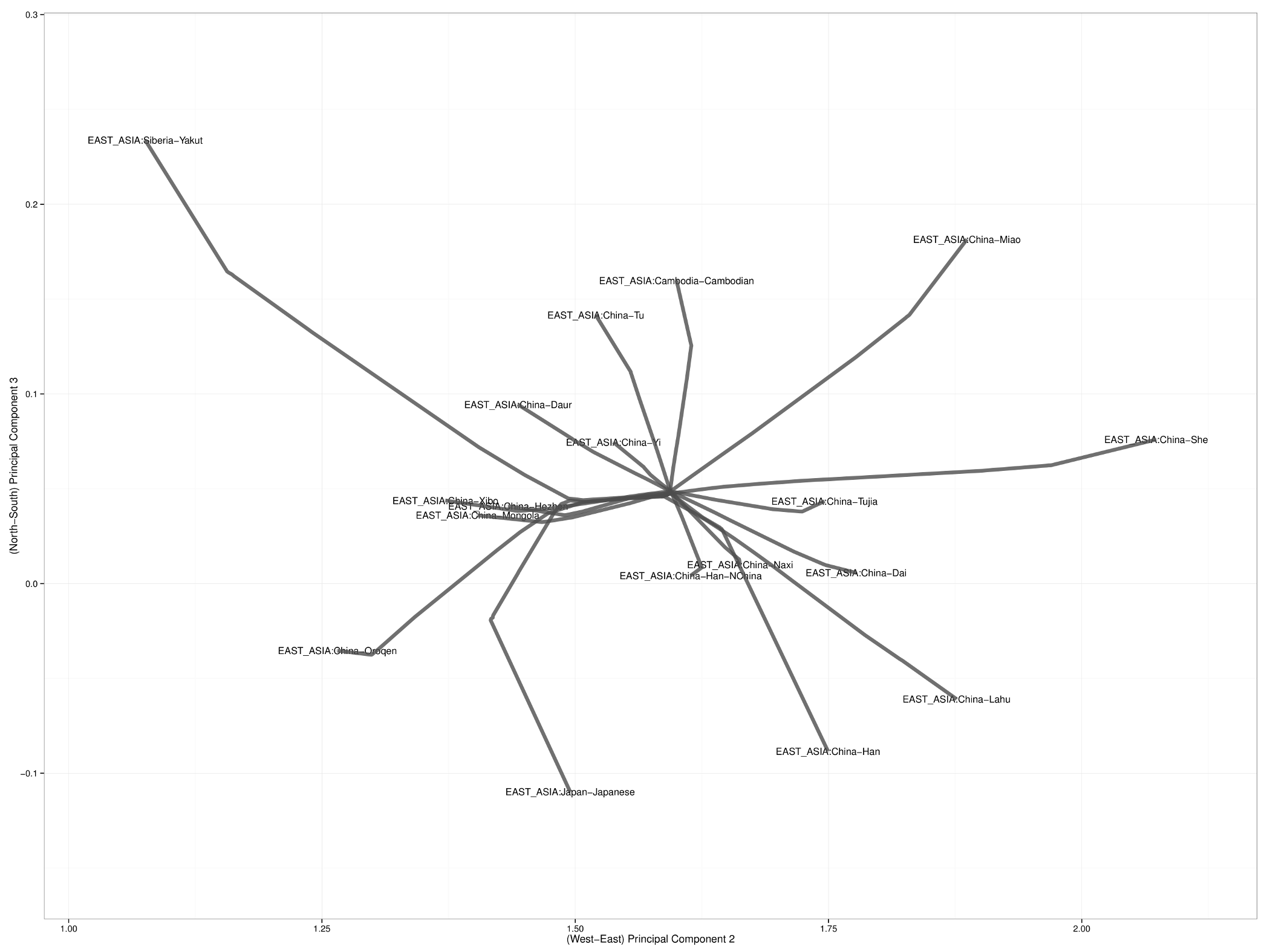}
\caption{\label{fig_convex_hgdp_eastasian}Magnified view of the convex clustering results for the HGDP data in East Asia.}
\end{center}
\end{figure}

\begin{figure}
\begin{center}
\includegraphics[width=7in]{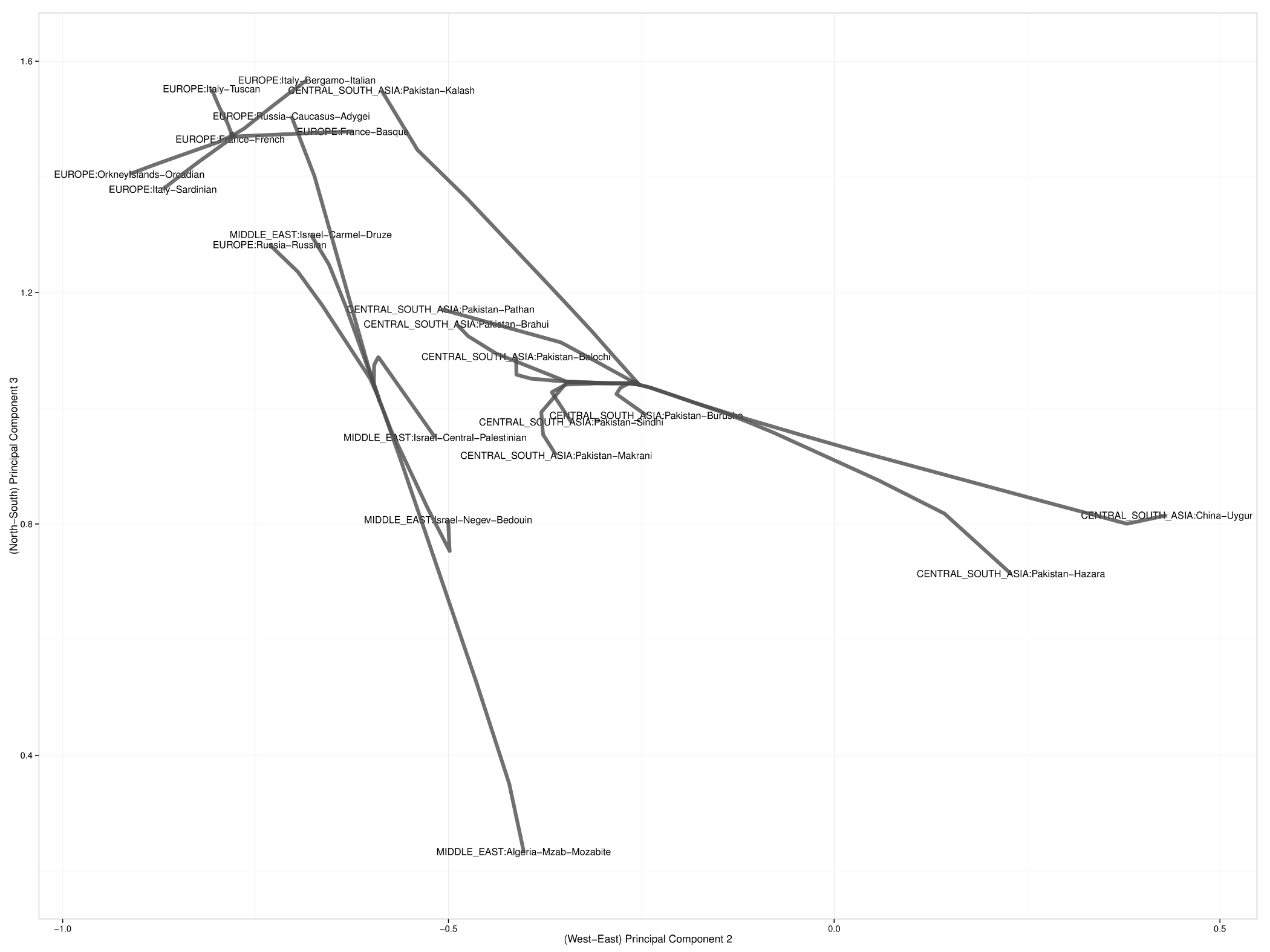}
\caption{\label{fig_convex_hgdp_europe_centralasia}Magnified view of the convex clustering results for the HGDP data in Europe and Central Asia.}
\end{center}
\end{figure}

Figures \ref{fig_convex_hgdp_eastasian} and \ref{fig_convex_hgdp_europe_centralasia} focus on related populations along the eastern and northern axes of East Asia, respectively. Most of the Chinese populations along the eastern axis appear to coalesce simultaneously. Some of the other populations along the northern border of China coalesce early. The Hezhen and Oroqen peoples reside predominantly in the Heilongjiang province of northeast China \citep{china90,china00}. These two populations cluster early with the inner Mongolians and the Xibo population, who occupy northeast China and the northwest region of Xinjiang province. Three distinct clusters of Middle Easterners, Central Asians, and Europeans occur along the northern axis. All European populations except for the Russian populations are grouped into a single cluster. The two Russian populations instead merge with a second cluster that includes three populations from Israel. The Mozabites, who coalesce late with this cluster, exhibit high frequencies of North African haplotypes as previously noted in the literature \citep{rosenberg06,coudray09}.  A third cluster within Central Asia unite Pakistani populations with Uygurs from China. Within this cluster, the Brahui, Balochi, and Makran populations of the Baluchistan province of northwestern Pakistan coalesce early with the Sindhi people of the Sindh province on the eastern border of Baluchistan. Later coalescing populations include the Hazara, Uygurs, and Kalash. The Hazaras of Pakistan and the Uygurs of China share common Mongolian and Turkic ancestry and some physical attributes \citep{qamar02,ablimit13}. A previous admixture analysis using high-density SNP data supports our observation that the Kalash people constitute a single distinct cluster, one of seven clusters separating all of the populations covered in the HGDP data \citep{rosenberg06}.

\begin{figure}
\begin{center}
\includegraphics[width=7in]{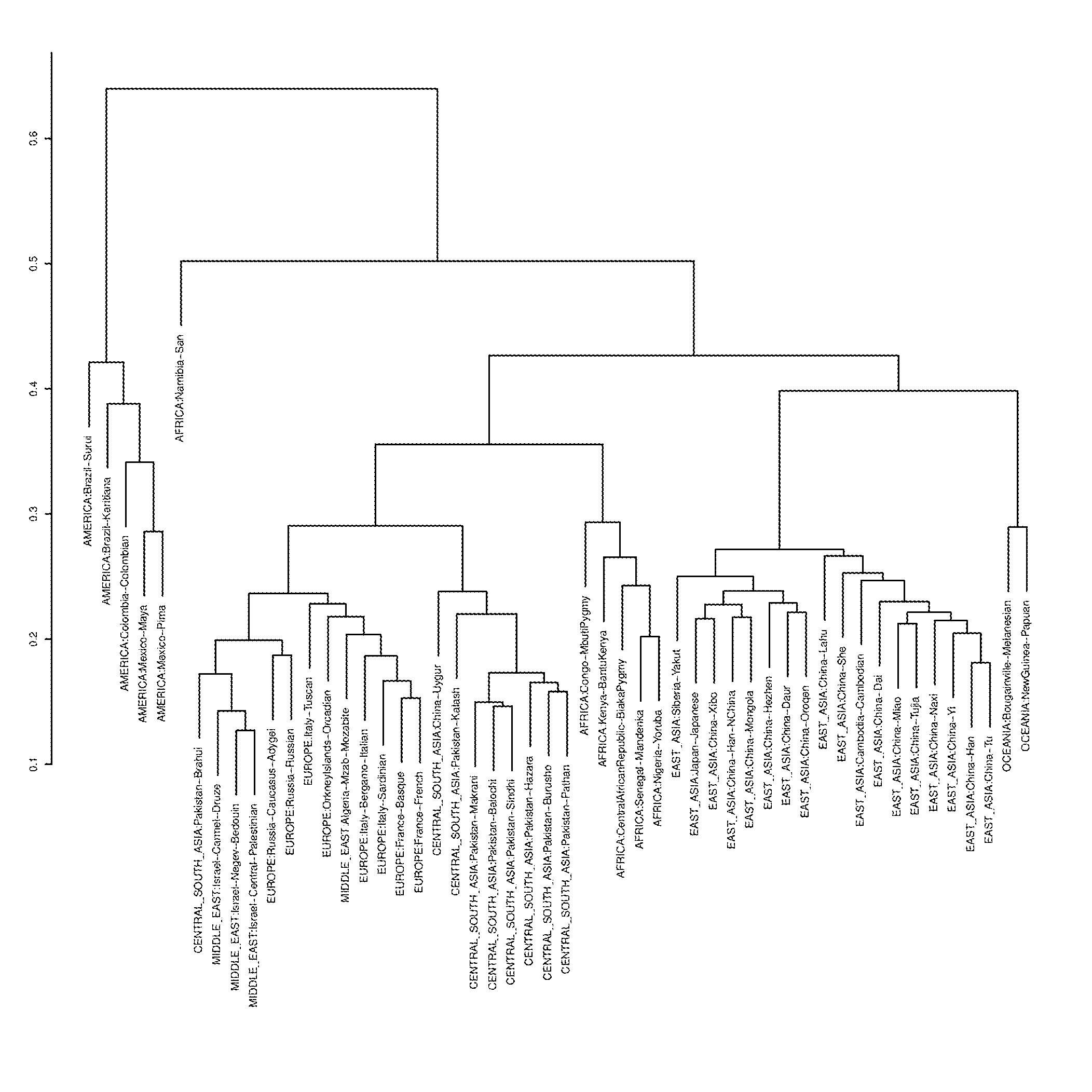}
\caption{\label{fig_hclust_hgdp}Hierarchical clustering of the 52 populations from the HGDP data.}
\end{center}
\end{figure}

The dendrogram in Figure \ref{fig_hclust_hgdp} presents the output of hierarchical clustering. Results are largely consistent across the two methods. For example, in East Asia both methods cluster the Chinese groups in the north of China with the Yakut of Siberia and the Japanese. The two populations from Papua New Guinea form their own cluster in both methods. However, some key differences occur. Convex clustering infers a closer genetic similarity between the Uygurs and the Hazara than suggested by hierarchical clustering. Based on evidence from admixture analyses \citep{rosenberg06} and convex clustering, one expects the Kalash to coalesce after the Uygur and other Central South Asian populations have coalesced. Although the non-Russian populations from Europe clustered in one group under both methods, hierarchical clustering included the Mozabites within this group as well, suggesting that they are more genetically related to Italians from Bergamo than Italians from Tuscany. This observation is inconsistent with our findings and previous work \citep{coudray09,rosenberg06}. Perhaps, the most puzzling discrepancy is the case where hierarchical clustering coalesces the African San population very late, only after all populations outside the Americas have coalesced.

\subsubsection{Population Structure of Europe}

We next investigate whether convex clustering can glean further insights into the population structure of Europe. The POPRES resource archives high-density genotypes generated on the Illumina 550k microarray platform \citep{nelson08}.  Version 2 of POPRES contains genotype and phenotype data on 4,077 subjects genotyped across 457,297 SNPs.  For this analysis, we include only non-admixed Europeans who report all four grandparents of the same ethnicity. This leaves 1,896 subjects. SNP data presents advantages and disadvantages compared to microsatellite data. Dense marker panels may be more sensitive to subtle differences driven by population events such as migration, expansion, and bottlenecks. Challenges include the lower information content of biallelic markers and the correlations between markers caused by linkage disequilibrium (LD). After considerable experimentation, we found that the leading principal components offered more insight into population structure than the raw genotypes themselves. We employed \eigenstrat~to extract the ten leading principal components from the genotype matrix. \eigenstrat~prunes SNPs in LD with $r^2$ exceeding a user-specified threshold \citep{price06}. In our case the threshold 0.8 discards all but 276,823 nearly independent SNPs. Our choice of the composite distance defined in equation (\ref{composite_distance}) places equal weight ($\alpha = 0.5$) on genetic distances and GPS distances between the capital cities of participants. To ease visualization, our figures display a maximum of 20 subjects from each ethnicity, for a total of 370 subjects. The computed convex clustering path is projected onto the first two principal components of the POPRES data; these components capture geographic east-west and north-south axes, respectively.

\begin{figure}
\begin{center}
\includegraphics[width=7in]{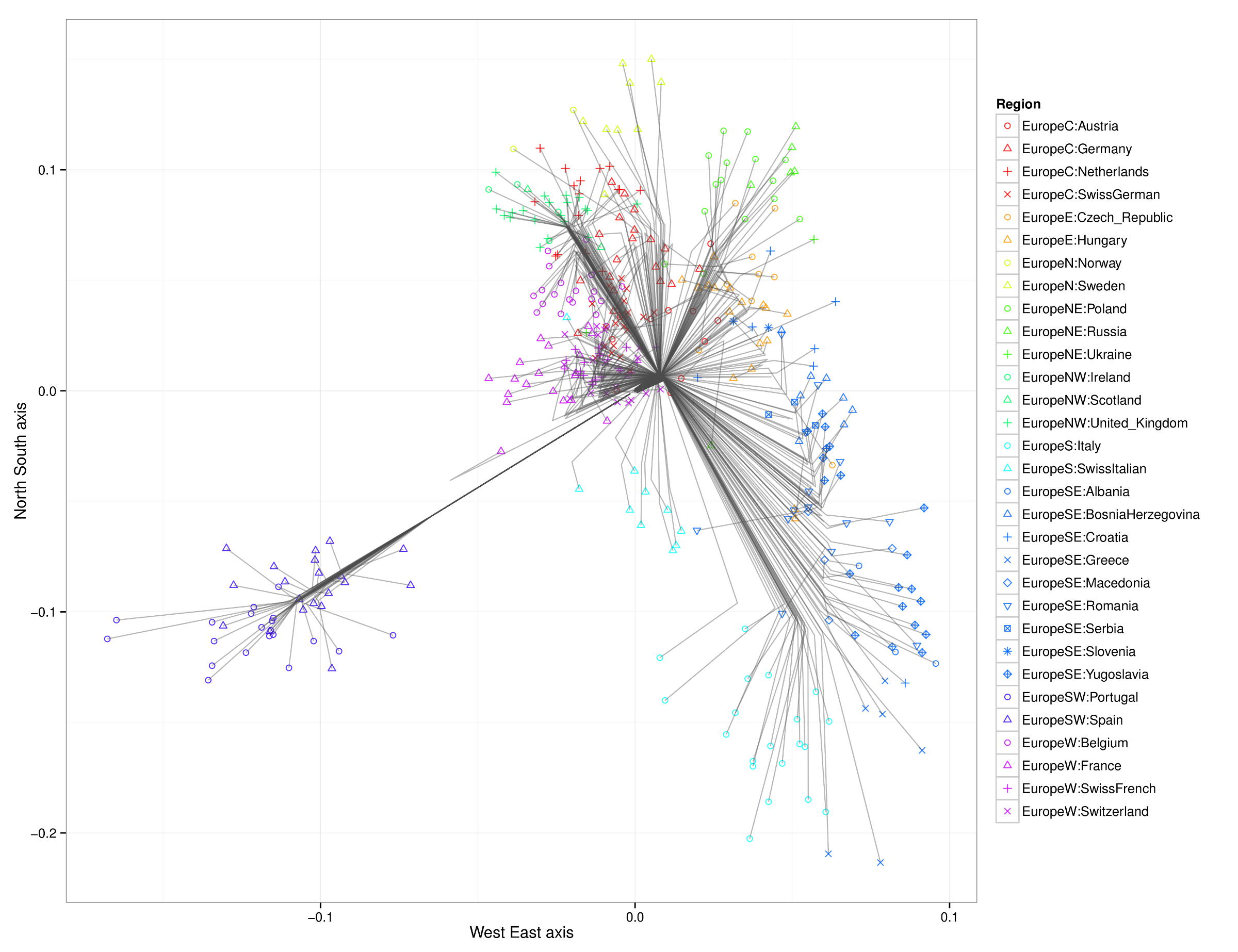}
\caption{\label{fig_convex_popres_phi00}Convex clustering of the European populations from the POPRES data using $\phi=0$ and $k=40$.}
\end{center}
\end{figure}

\begin{figure}
\begin{center}
\includegraphics[width=7in]{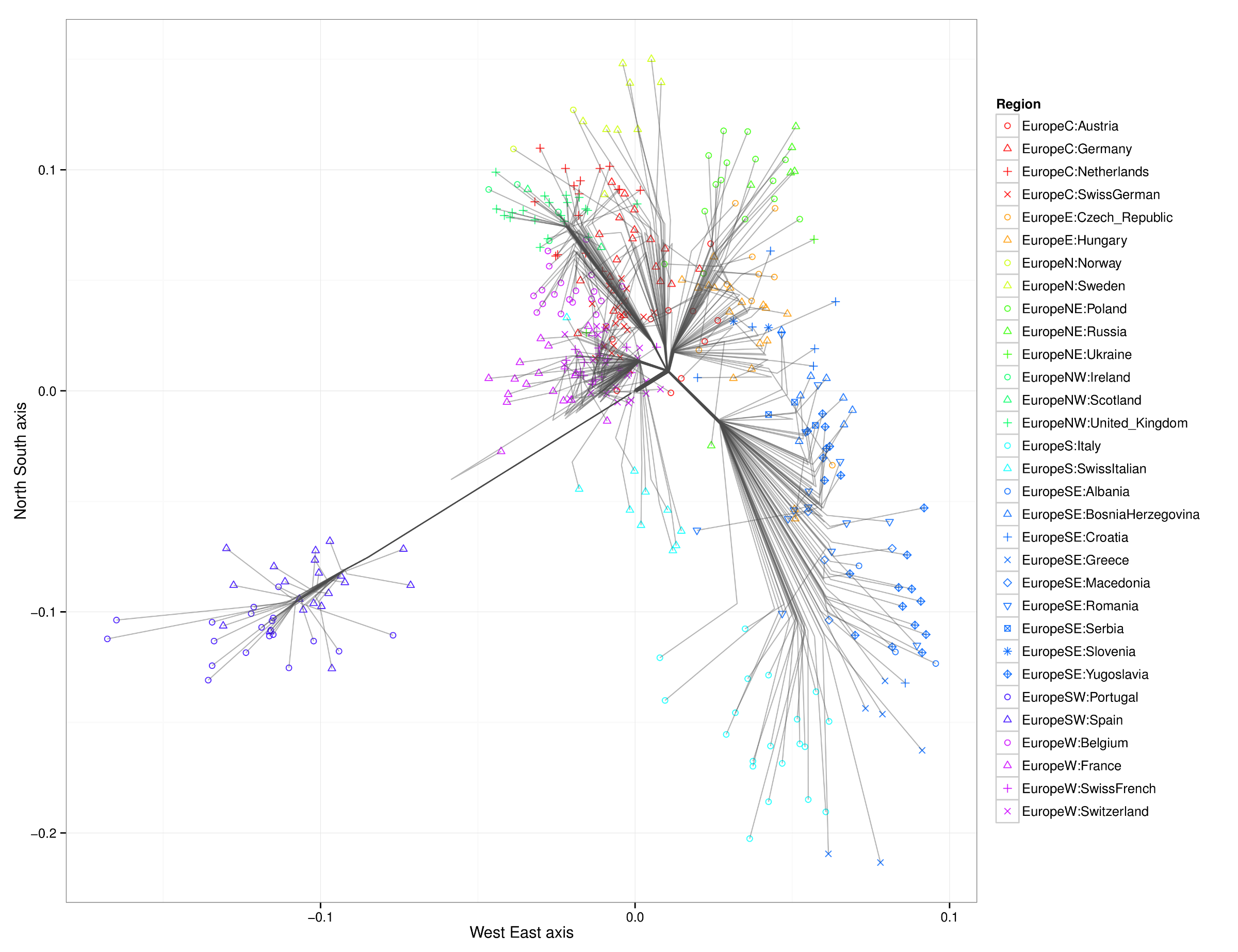}
\caption{\label{fig_convex_popres_phi10}Convex clustering of the European populations from the POPRES data using $\phi=10$ and $k=40$.}
\end{center}
\end{figure}

In the Iris and the HGDP datasets, the number of nearest neighbors $k$ was more critical in resolving cluster evolution than the tuning constant $\phi$. In the European POPRES data, where inter-class differences are more subtle, increasing $\phi$ can be critical in resolving details for $k$ large. Figure \ref{fig_convex_popres_phi00} depicts a clustering path with $k=40$ neighbors and $\phi=0$. Increasing $\phi$ to 10 gives a similar clustering pattern, except that each of the major trunks coalesce before converging to the origin. Thus, Figure \ref {fig_convex_popres_phi10} shows several major clusters connected by five major trunks. Spain and Portugal constitute a major cluster in the southwest trunk. The southeast trunk includes Italy and southeast Europe; these populations eventually merge into a single cluster. The northeast trunk defines a cluster that includes Poland, Russia, Ukraine, the Czech Republic, Hungary, and Slovenia. Norway, Sweden, and Germany cluster along the northern trunk, and the British Isles merge with Belgium and the Netherlands to form the northwest trunk. A large cluster comprising France and the Swiss linguistic groups (French, German, and Italian) constitute the western trunk. Replotting the clustering path with $\phi=1$ and $k=3$ shows Norway and Sweden breaking away from Germany and forming their own disjoint cluster. France breaks away from the Swiss groups to form its own disjoint cluster. Along the south trunk, Italy now separates from southeast Europe and eventually clusters with the Swiss-Italians.  

\begin{figure}
\begin{center}
\includegraphics[width=7in]{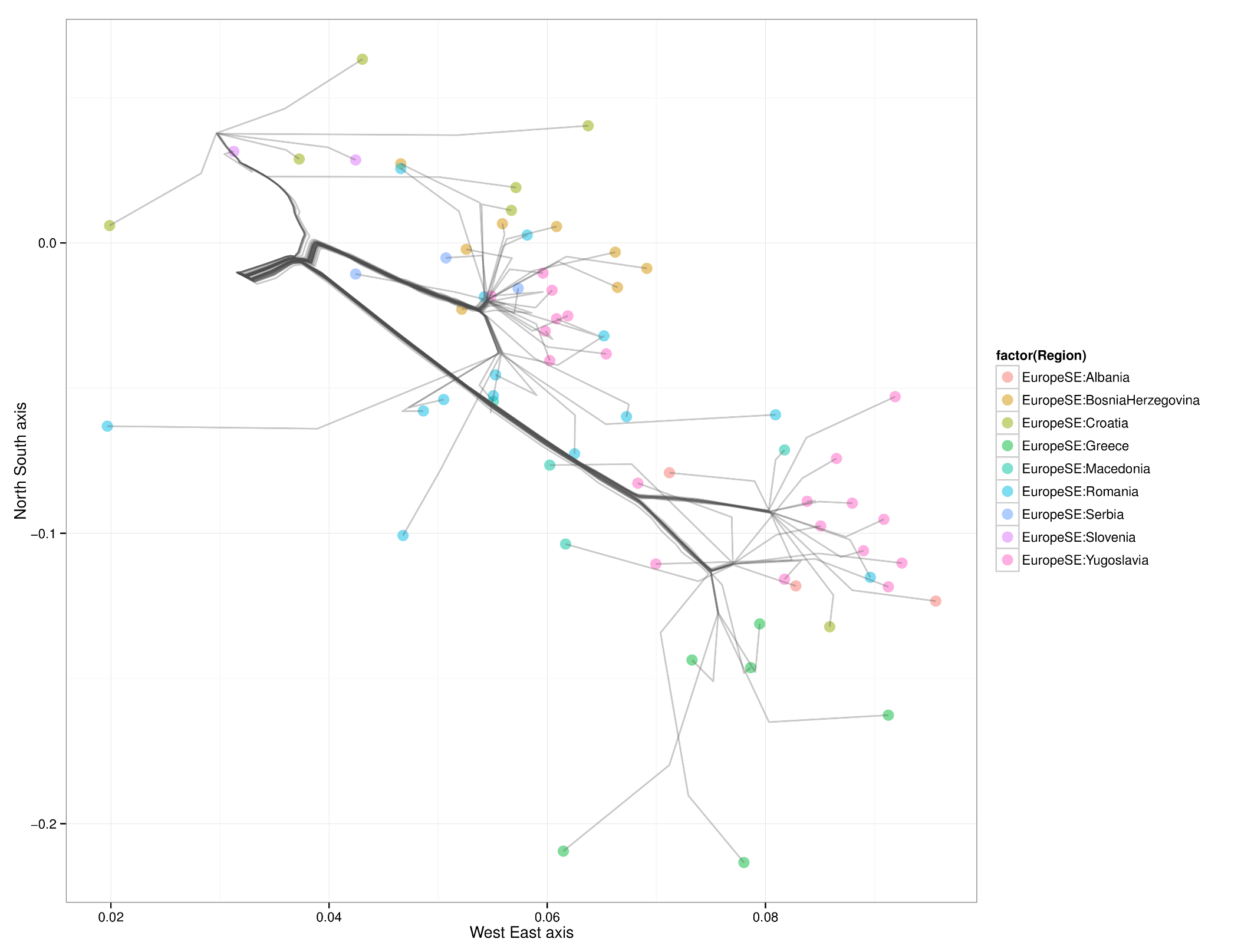}
\caption{\label{fig_convex_europeSE}Magnified view of results from convex clustering of Southeast Europe.}
\end{center}
\end{figure}

\begin{figure}
\begin{center}
\includegraphics[width=7in]{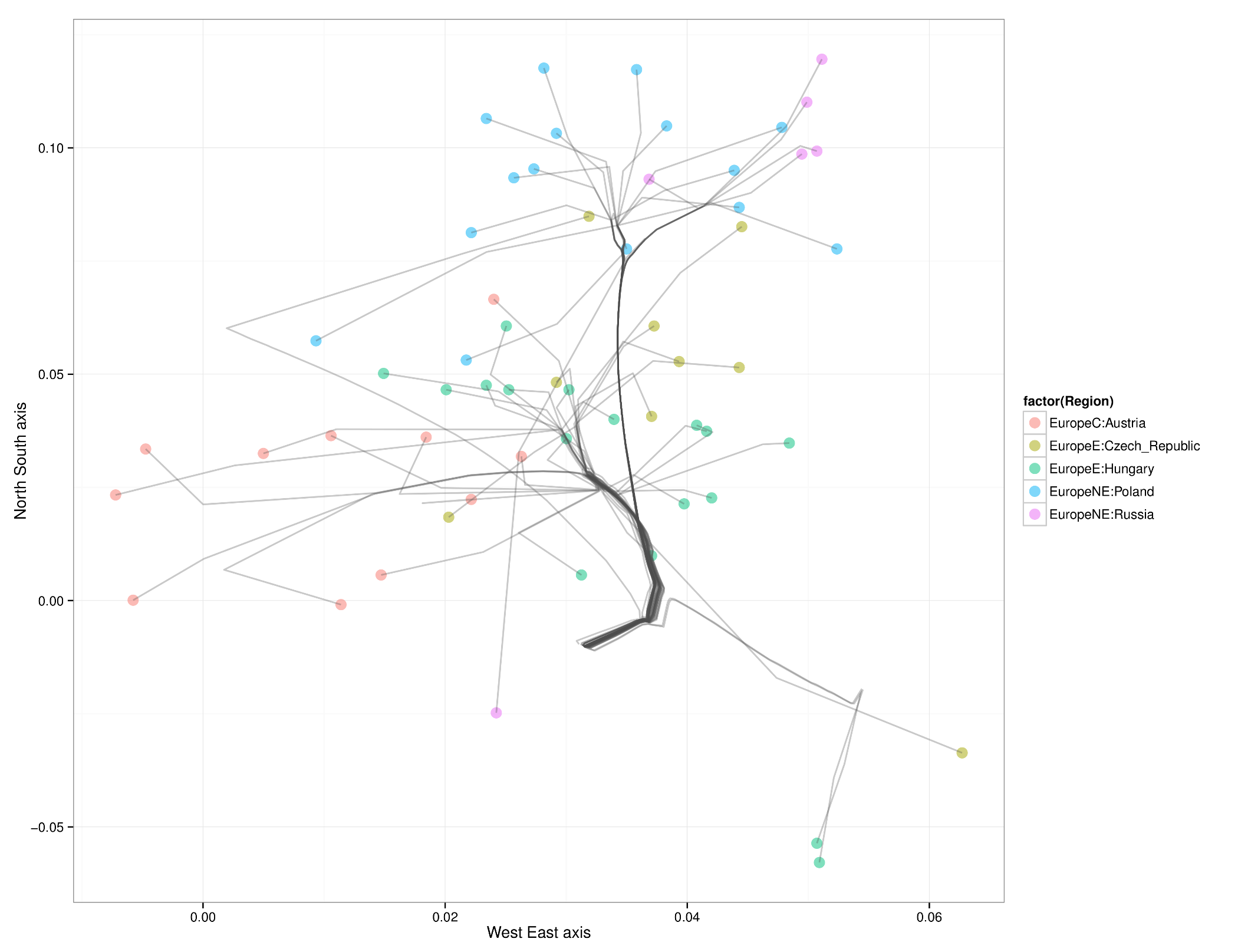}
\caption{\label{fig_convex_europeNE}Magnified view of results from convex clustering of Northeast Europe.}
\end{center}
\end{figure}

Figure \ref{fig_convex_europeSE} depicts the clustering path of southeast Europe, where West Slavic languages predominate. Here Greece first coalesces with Macedonia, a Slavic population bordering Greece on the north.  A cluster comprising Bosnia-Herzegovina and Serbia merges with Romania, before merging into the primary trunk of southeast Europe. Finally at the northern end of the trunk, a cluster formed by Croatia and Slovenia form its own cluster. The groups in the Bosnia-Herzegovina cluster and the Macedonian cluster are consistent with local history. Poland and Russia cluster in the northern most branch of the northeast trunk (Figure \ref{fig_convex_europeNE}). The Czech-Republic, Austria, and Hungary define a distinct cluster along the southern branch. Given that Austria conquered Hungary in 1699 and established rule over Bohemia (the predecessor to modern Czechs) as early as 1526, these results are not surprising.

\begin{figure}
\begin{center}
\includegraphics[width=3in]{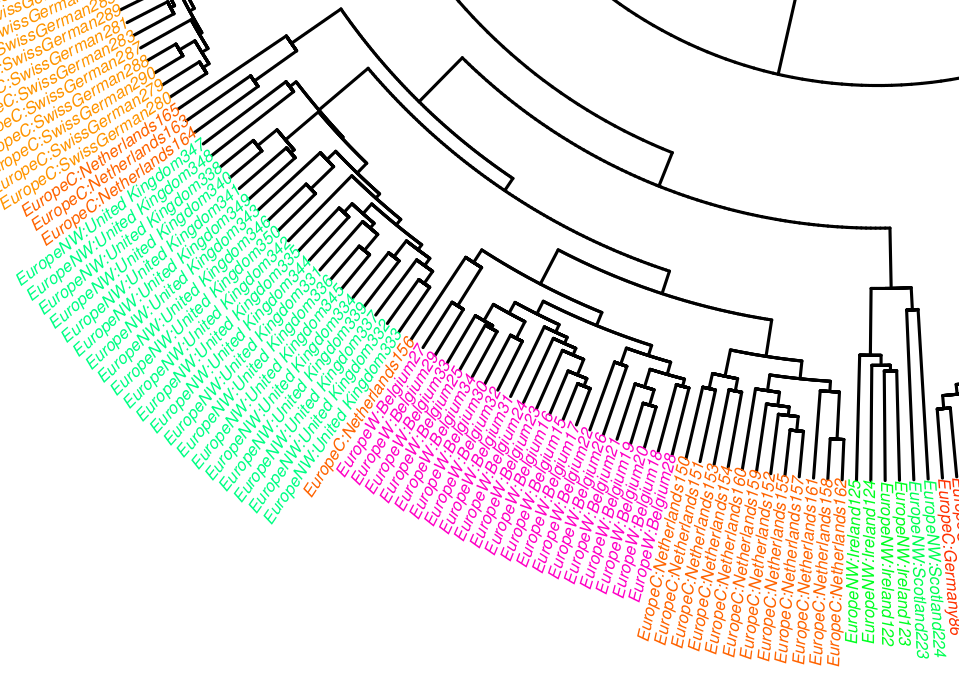}
\caption{\label{fig_hclust_british} UPGMA dendrogram showing genetic relationships among populations in and near the British Isles.}
\end{center}
\end{figure}

In the POPRES data, convex clustering and hierarchical clustering occasionally disagree. For example, hierarchical clustering merges the Netherlands and Belgium with Britain before it merges Britain with Ireland and Scotland (Figure \ref{fig_hclust_british}). In light of the geography and history of Britain, it is reasonable to expect Britain to first merge with Scotland and Ireland. Convex clustering produces this intuitive cluster, which eventually merges with the neighboring cluster comprising Belgium and the Netherlands.  Owing to a few outliers, the greedy nature of hierarchical clustering appears to force a spurious coalescence, which cannot be repaired until later. Another discrepancy occurs in clustering the Swiss linguistic groups. Convex clustering first groups the Swiss-German, Swiss-French, and Swiss-Italian into a single Swiss cluster. Hierarchical clustering groups France with this cluster. At the next higher level, hierarchical clustering fails to cluster Italy with the Swiss and instead merges it with Greece and populations from the former Yugoslavia. Convex clustering, on the other hand,  produces a more intuitive path, merging Italy with the Swiss before joining both to the southeast European trunk.

\subsection{Inferring Cancer Subtypes}

\begin{figure}
\begin{center}
\includegraphics[width=7in]{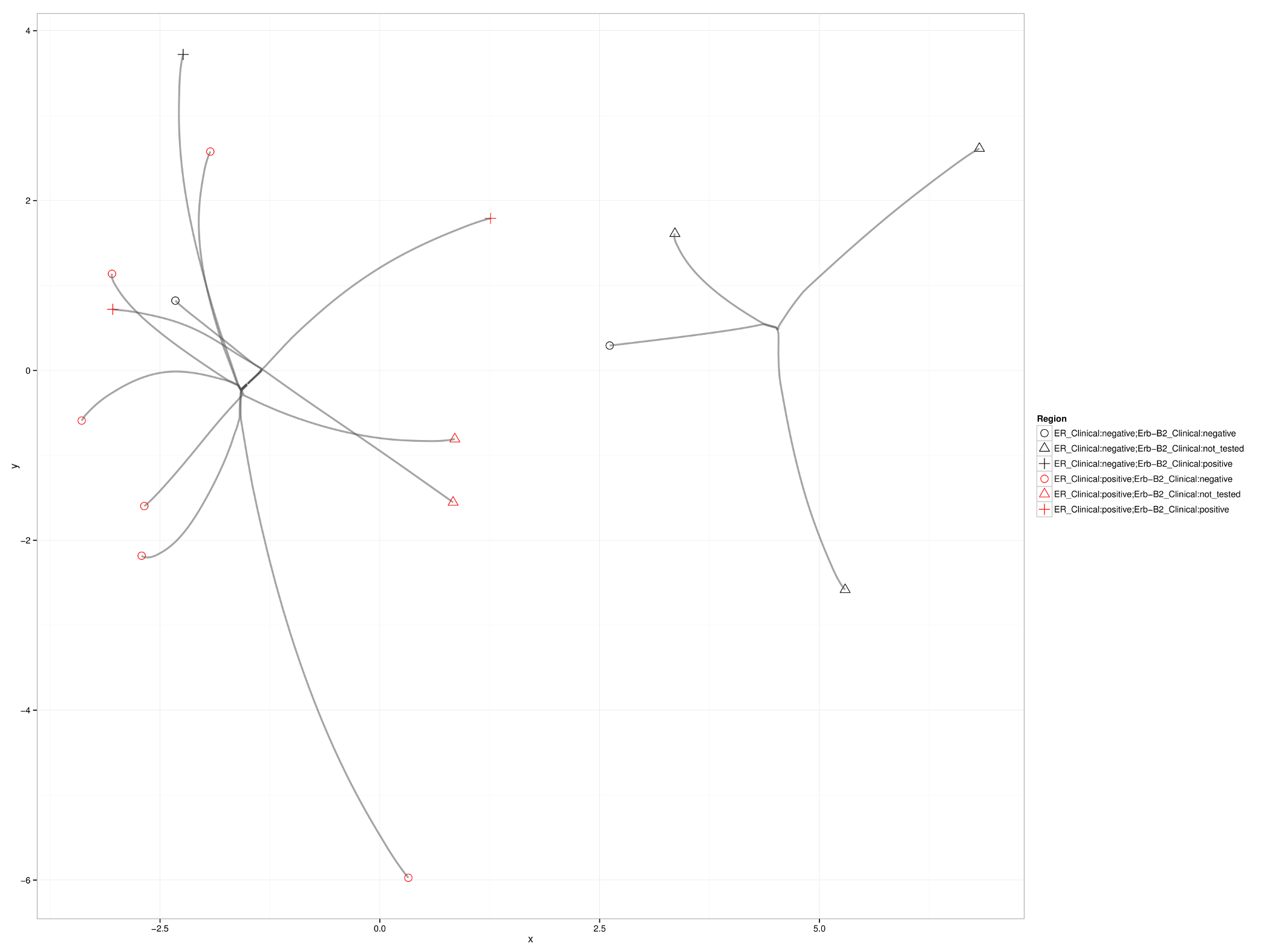}
\caption{\label{fig_convex_expression}Convex clustering of the breast cancer samples. Points on the plot indicate data vectors projected onto the first and third principal components (PCs) of the sample. Lines trace the cluster centers as they traverse the regularization path.}
\end{center}
\end{figure}

\begin{figure}
\begin{center}
\includegraphics[width=7in]{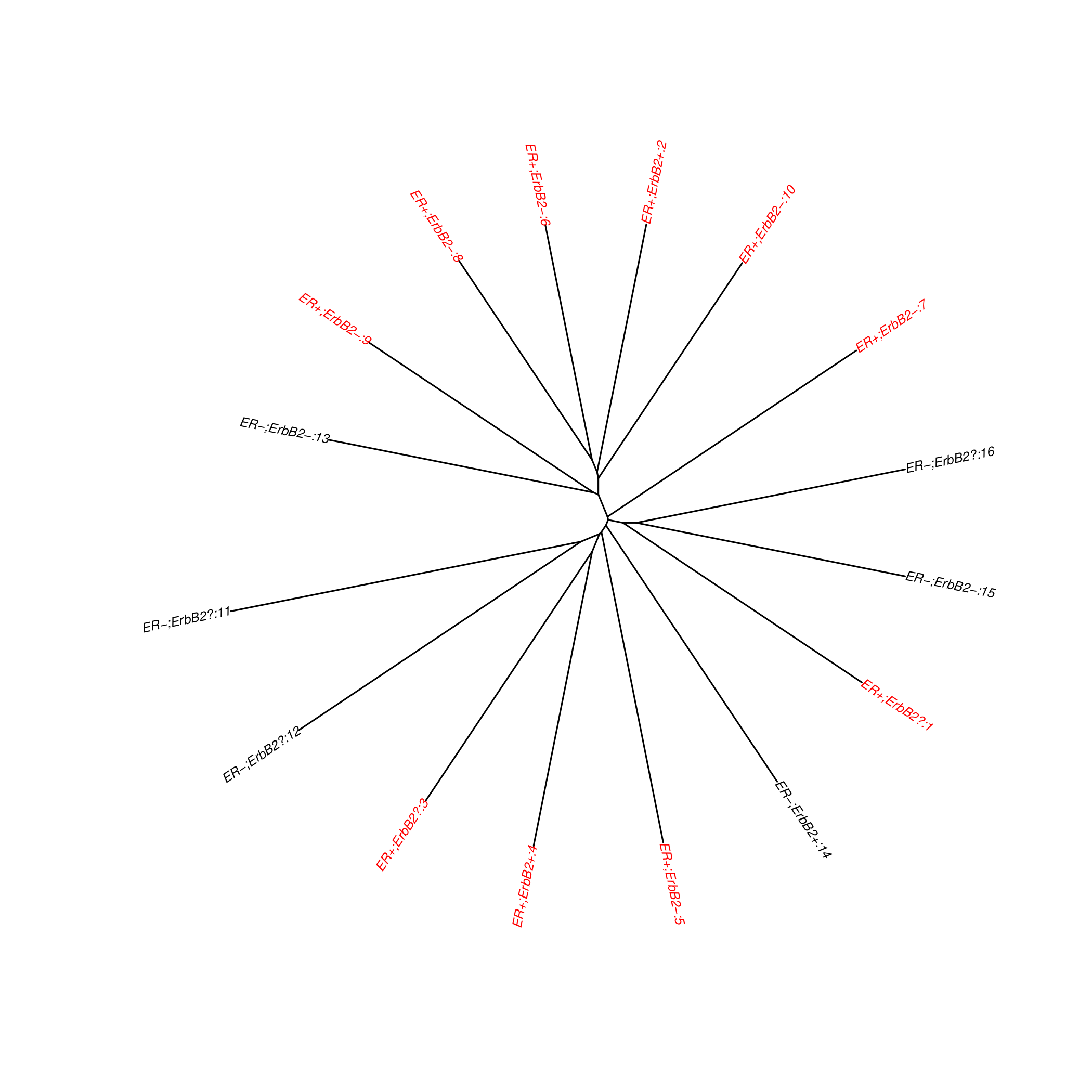}
\caption{\label{fig_hclust_expression}Average linkage hierarchical clustering of the breast cancer samples.}

\end{center}
\end{figure}

It is well accepted that cancers of a given tissue often fall into different subtypes.  In breast cancer for instance, patients with tumors that are estrogen receptor (ER) negative are less responsive to hormone based treatment than those possessing active estrogen receptors (ER) \citep{rochefort03}. High-throughput platforms such as gene-expression microarrays and RNA-Seq have enabled researchers to classify cancer patients based on their molecular phenotypes. Hierarchical clustering by \cite{perou00} established five gene-expression profiles across 9216 genes in 84 breast-cancer patients. Among the 84 patients, only 16 also had a clinical assessment of ER status.  It is natural to ask whether convex clustering can discriminate between the two ER tumor types. Under the tuning constants $\phi=.5$ and $k=1$, convex clustering recovers two distinct clusters. Figure \ref{fig_convex_expression} projects the cluster centers along the cluster path on the first and third principal components of the original data. The left and right clusters correspond roughly to ER positive and ER negative tumors, respectively.  Two ER negative tumors appear as outliers in the ER positive cluster. Within the ER positive cluster, two ER positive samples with ER-B2 positive status cluster before merging with the remaining samples. Figure \ref{fig_hclust_expression} depicts an unrooted tree derived from hierarchical clustering. In comparison to convex clustering, the cluster definitions are vague, with most clusters containing two or three members. This could be an artifact of the hard binary choices imposed by hierarchical clustering. The two ER-B2 positive samples that clustered together in convex clustering appear in distant clusters under hierarchical clustering.

\subsection{Run-time benchmarks}

For a dataset with a large number of attributes, parallelization can substantially reduce run times. \convexcluster~includes code written in OpenCL, a language designed to run on many-core devices such as GPUs. For each of the three genetic analyses presented above, we recorded the total run-time along the entire regularization path using standard C++ code for the CPU and OpenCL code for the GPU. For the sake of comparison, we also recorded run-times for \clusterpath~on the same datasets and weighting schemes. Table \ref{table_gpu} records the average run time to minimize the objective function averaged over all values of the regularization parameter. We chose this strategy because \clusterpath~does not allow users to pre-specify a grid of regularization values.  The bottom line is that \convexcluster~required only 16\%, 47\%, and 75\% of the time required by \clusterpath~to fit the HGDP, POPRES, and breast cancer datasets respectively. When a GPU is available, further improvements can potentially be realized. On an nVidia C2050 GPU, \convexcluster~enjoys speed improvements of 4.6 and 5.5 fold over the CPU version for the HGDP and breast cancer examples. In contrast, on the POPRES example, the GPU version is actually 3.5 fold slower than the CPU version. GPU programs suffer when a significant amount of time is spent transferring data to and from the GPUs. In its current form, \convexcluster~reads the updated matrix $\bU$ from the GPUs at each point on the $\mu$-regularization path before saving the data to disk. This large I/O overhead can overwhelm gains from parallelization for low-dimensional datasets such as the POPRES data. In general, 
GPU implementations of standard algorithms require a high degree of parallelization, limited data transfers between the master CPU and the slave GPUs, and maximal synchrony of the GPUs.
Depending on the nature of the clustering data, \convexcluster~satisfies these requirements.

\section{Discussion}

The literature on cluster analysis is enormous. Each clustering method has advantages in either simplicity, speed, reliability, interpretability, or scalability. If the number of clusters is known in advance, then $k$-means clustering is usually preferred. In convex clustering one can often achieve a predetermined number of clusters by varying the number of nearest neighbors and following the solution path to its final destination. Alternatively, if the underlying graph is fully connected, then one can follow the solution path until $k$ clusters appear. The downside of $k$-means clustering is that it offers no insight into cluster similarity. If the goal in clustering is to obtain a snapshot of the relationships among observed data points at different levels of granularity, the choices are limited, and most biologists opt for hierarchical clustering. Hierarchical clustering is notable for its speed and visual appeal. Balanced against these assets is its sensitivity to poor starting values and outliers. Our perturbations of the Iris data demonstrate the latter weakness. Convex clustering occupies an enviable middle ground between $k$-means clustering and hierarchical clustering. Our extensive exploration of the HGDP and POPRES datasets showcase the subtle solutions paths of convex clustering. These paths offer considerable insights into population history and correct some of the greedy mistakes of hierarchical clustering.

Given the novelty of convex clustering \citep{lindsten11}, it is hardly surprising that only a single previous program, \clusterpath, implements it \citep{hocking11}. \convexcluster~and 
\clusterpath~perform similarly on modest problems such as the Iris data. Unfortunately, on large 
datasets such as the HGDP data, \clusterpath~depletes all available memory and fails. Furthermore, \clusterpath~lacks two features that work to the advantage of convex clustering. First, it does not support disconnected graphs defined by sparse weights. In our breast cancer example, clustering with disconnected
graphs reveals fine-grained details. Second, \clusterpath~does not allow for missing entries in the data matrix. The current paper documents \convexcluster's ability to scale realistically to dimensions typical of modern genomic data. A combination of careful algorithmic development and exploitation of modern many-core chipsets lies behind \convexcluster. The proximal distance algorithm propelling \convexcluster~separates parameters and enables massive parallelization. OpenCL made it relatively easy to implement parallel versions of our original serial code. Further speedups are possible. For instance, \convexcluster~spends an inordinate amount of execution time moving matrices over relatively slow I/O channels in preparation for plotting. One could easily project the data to principal components on each GPU itself prior to data transfer. More recent ATI or nVidia GPUs should improve the speedups on high-dimensional data mentioned here.

Convex clustering also shows promise as a building block for more sophisticated exploratory tools in computational biology. In a companion paper \cite{ChiAllBar2014} introduce a convex formulation of the biclustering problem. In biclustering one seeks to cluster both observations and features simultaneously in a data matrix. Cancer subtype discovery can be formulated as a biclustering problem in which gene expression data is partitioned into a checkerboard-like pattern highlighting the associations between groups of patients and the groups of genes that distinguish them. To bicluster a data matrix, hierarchical clustering can be applied independently to the rows and columns of the matrix. Convex biclustering produces more stable biclusterings while retaining the interpretability of hierarchical biclustering. Convex biclustering requires repeatedly solving convex clustering subproblems.

The field of cluster analysis is crowded with so many competing methods that it would foolish to conclude that convex clustering is uniformly superior. Our goal of illustrating the versatility of convex clustering is more modest. The reflex reaction of most biologists is to employ hierarchical or $k$-means clustering. We suggest that biologists take a second look. Convex clustering's ability to reliably deliver an entire solution path is compelling. The insights discussed here will enhance the careful exploration of many big datasets. The present algorithm, and indeed the present formulation of convex clustering, are unlikely to be the last words on the subject. We encourage other computational biologists and statisticians to refine these promising tools. \convexcluster~can be freely downloaded from the UCLA Human Genetics web site at http://www.genetics.ucla.edu/software/ for analysis and comparison purposes.     



\bibliographystyle{natbib}

\bibliography{ChenCluster}

\clearpage

%
%

%

\begin{table}[!t]
\begin{center}
\caption{Error rates as a function of noise in the Iris Data\label{table_simulation}}
\vspace{0.05in}
\begin{tabular}{|c|c|c|c|c|}
\hline
Noise level & \hclust & \multicolumn{3}{c|}{\convexcluster} \\
	$c$ & UPGMA & \multicolumn{1}{c}{k=5} & \multicolumn{1}{c}{k=10} 
& \multicolumn{1}{c|}{k=15}	\\
\hline
0.02 & 0.193 & 0.105 & 0.095 & 0.102\\
0.04 & 0.194 & 0.108 & 0.107 & 0.105\\
0.06 & 0.197 & 0.107 & 0.104 & 0.105\\
0.08 & 0.216 & 0.111 & 0.107 & 0.126\\
0.10 & 0.226 & 0.120 & 0.121 & 0.134\\
\hline
\end{tabular}
\end{center}
\end{table}

\begin{table}[!t]
\begin{center}
\caption{Error rates as a function of missingness in the Iris Data\label{table_missingness}}
\vspace{0.05in}
\begin{tabular}{|c|c|c|c|c|}
\hline
Proportion of rows with & \hclust & \multicolumn{3}{c|}{\convexcluster} \\
a missing attribute	$c$ & UPGMA & \multicolumn{1}{c}{k=5} & \multicolumn{1}{c}{k=10} 
& \multicolumn{1}{c|}{k=15}	\\
\hline
0.25 & 0.210 & 0.109 & 0.115 & 0.127\\
0.50 & 0.205 & 0.127 & 0.137 & 0.133\\
0.75 & 0.228 & 0.147 & 0.148 & 0.141\\
1.00 & 0.228 & 0.153 & 0.181 & 0.146\\
\hline
\end{tabular}
\end{center}
\end{table}

\begin{table}[!t]
\begin{center}
\caption{Average runtimes in seconds for different analyses \label{table_gpu}}
\vspace{0.05in}
\begin{tabular}{|l|r|r|r|r|r|}
\hline
Analysis &  Datapoints & Variables & \clusterpath~  &  \multicolumn{2}{c}{\convexcluster} \\
& &   &  & CPU &  GPU \\
\hline
HGDP & 52  & 4,682 & 8.67 & 1.46 &  .32 \\
POPRES & 370 & 10 & 2.53 & 1.21 & 4.29 \\
Breast Cancer data & 16 & 9,216 & 3.14 & 2.37 &  .43 \\
\hline
\end{tabular}
\end{center}
\end{table}

\end{document}